\documentclass[aps,prb,twocolumn,showpacs]{revtex4}
\usepackage[utf8]{inputenc}
\usepackage[english]{babel}
\usepackage{xcolor}
\usepackage{braket}
\usepackage{bm,amsmath}
\usepackage{amssymb}
\usepackage{graphicx}
\usepackage[active]{srcltx}
\usepackage{subfigure}
\newcommand{\pda}{^{\phantom\dagger}}
\renewcommand{\vec}{\textbf}
\renewcommand{\Re}{\mathrm{Re}}

\begin{document}
     \title{Coulomb-assisted cavity feeding in the non-resonant optical emission
from a quantum~dot}

    \author{Matthias Florian$^1$, Paul Gartner$^{1,2}$, Alexander Steinhoff$^1$, Christopher Gies$^1$
and Frank Jahnke$^1$}
    \affiliation{$^1$ Institute for Theoretical Physics, University of Bremen,
28334 Bremen, Germany}
    \affiliation{$^2$ National Institute of Materials Physics,
Bucharest-Magurele, Romania}
    \date{\today}
    \begin{abstract}
        Recent experiments have demonstrated that for a quantum dot in an
optical resonator off-resonant cavity mode emission can occur even for detunings
of the order of 10 meV. We show that Coulomb mediated Auger processes based on additional
carriers in delocalized states can facilitate this far off-resonant emission. 
Using a novel theoretical approach for a non-perturbative treatment 
of the Auger-assisted quantum-dot carrier recombination, we present numerical
calculations of the far off-resonant cavity feeding rate and cavity mean photon number confirming
efficient coupling at higher densities of carriers in the delocalized
states. In comparison to fast Auger-like intraband scattering processes, 
we find a reduced overall efficiency of Coulomb-mediated interband transitions due the
required electron-hole correlations for the recombination processes.
    \end{abstract}

    \maketitle

    Semiconductor quantum-dot (QD) microcavity devices offer many applications
of strong current interest, such as lasers with improved emission
properties and integrated sources of indistinguishable and entangled
photons~\cite{alferov_nobel_2001, bimberg_quantum_1999, strauf_single_2011,
michler_single_2003, michler_single_2009}. In contrast to atomic-like isolated emitters,
QDs exhibit an interesting peculiarity: even if the QD emission lines are significantly detuned
from the cavity resonance, photons can be emitted into the cavity mode. As a result of intense
experimental and theoretical investigations, different mechanisms are discussed in the literature. 
Firstly, the interaction of QD excitons with acoustic phonons has been shown 
to provide efficient off-resonant cavity feeding for an energetic mismatch of few 
meV~\cite{press_photon_2007, tarel_influence_2008, ates_non-resonant_2009, hohenester_phonon-assisted_2009,majumdar_phonon_2011}. 
Secondly, when a QD can accommodate many single-particle bound states,
the number of carrier configurations becomes quite large, and their Coulomb
interaction results in a broad quasi-continuum of multiexcitonic transitions.
Provided that these multi-exciton states are excited, their
overlap with the cavity mode ~\cite{dekel_multiexciton_1998,
winger_explanation_2009, laucht_temporal_2010} allows for a Purcell-enhanced
photon production at larger detunings ($\sim 10$ meV away from the single
exciton line). Also the role of the interaction with the wetting layer (WL) in the
formation of this multiexcitonic spectral background was recognized~\cite{winger_explanation_2009} and Coulomb hybridization of QD bound
states with the WL continuum was demonstrated~\cite{karrai_hybridization_2004, chauvin_controlling_2009}.

In this letter, we quantify the role of Coulomb interaction with WL carriers
for the off-resonant coupling of QD transitions to a cavity mode. Specifically, 
carriers in the WL can act as a thermal bath which compensate for the energy
mismatch via Auger-like processes. We point out that this is an
alternative mechanism to the Coulomb configuration interaction between 
carriers~\cite{winger_explanation_2009}, as we consider the role of the Coulomb
interaction not in the spectrum but in the dynamics of the exciton
recombination. Its effect in opening a kinetic channel is present even for QDs
hosting very few confined states, and as a proof of principle, we evaluate it
for a QD with single electron and hole levels.

We find that optical interband transitions assisted by WL carriers via
Coulomb interaction can provide off-resonant coupling, however, with smaller
rates in comparison to ultrafast intraband Coulomb relaxation
processes, which are known to allow carriers to efficiently bridge several tens
of meV~\cite{bockelmann_electron_1992, efros_breaking_1995,
uskov_auger_1997, nielsen_many-body_2004}. We trace back the reason for the
difference to electron-hole correlation effects.

To describe the system, we start from the Hamiltonian containing the Jaynes-Cummings (JC) part
for the interaction of the QD exciton with photons of the cavity mode as well as
the Coulomb part for the interaction between QD and WL states.
    Two different techniques are used to formally eliminate either (i) the
exciton-photon or (ii) the Coulomb interaction part from the Hamiltonian.   
Specifically, the Schrieffer-Wolff transformation~\cite{mahan_many_2000,
hohenester_cavity_2010} is used that amounts to a perturbative diagonalization 
of the JC interaction part.    As an alternative second method, we exploit a
novel approach, involving a unitary transform which is related to the one used
in the independent boson model (IBM) for the treatment of the carrier-phonon
interaction, but is now applied to the fermionic bath of WL carriers.

    The cavity-QD system in the presence of the continuum of WL states is
described by the Hamiltonian~(see Appendix~\ref{sec:exc_WL}, $\hbar=1$)
    \begin{align}
        \begin{split}
            H =&\,\varepsilon_X X^\dagger X + \omega b^\dagger b + g(b^\dagger X
+ b X^\dagger)\\
            &+(1 - X^\dagger X)h_0 + X^\dagger X h_X\,, \\
            h_0 =& \sum_{\lambda,\vec k} \varepsilon^{\lambda}_{\vec k}
\lambda^\dagger_{\vec k} \lambda\pda_{\vec k}\,,
            \quad\quad h_X = h_0 + W \, ,
            \end{split}
            \label{eq:hamiltonian}
    \end{align}
    where $X$ and $X^\dagger$ are the exciton annihilation and creation
operators corresponding to the QD-bound electron and hole $s$-states, $b$,
$b^\dagger$ are the photon operators, $\varepsilon_X$ and $\omega$ are the
exciton and cavity mode energy, $\lambda^{\pda}_{\vec k}$,
$\lambda^\dagger_{\vec k}$ are the fermionic operators referring to the
WL $\vec k$-states (including the spin implicitly) for electrons and holes
($\lambda = e,h$) with the energies $\varepsilon^{\lambda}_{\vec k}$, and $g$
is the JC coupling strength. 
    In the presence of the exciton, the WL free Hamiltonian $h_0$ is modified by
the interaction of the WL carriers with the electron-hole pair, described by the
operator
    \begin{align}
        W &= \sum_{\lambda,\vec k,\vec k'} \,W^{\lambda}_{\vec k\vec k'}
\lambda^\dagger_{\vec k}\lambda^{\pda}_{\vec k'} \,,
    \end{align}
    where
    \begin{align}
        W^e_{\vec k \vec k'} &= W^{ee}_{s\vec k,\vec k's}-W^{he}_{s\vec k,\vec
k's} - W^{ee}_{s\vec k,s\vec k'} \, ,
        \label{eq:we} \\
        W^h_{\vec k \vec k'} &= W^{hh}_{s\vec k,\vec k's}-W^{eh}_{s\vec k,\vec
k's} - W^{hh}_{s\vec k,s\vec k'} \,
        \label{eq:wh} \, .
    \end{align}
    The interaction matrix elements
    $W^{\lambda\lambda'}_{ij,kl} = \frac{1}{A}\sum_{\vec q}w_{\vec
q}\bra{\varphi^{\lambda}_i} e^{i\vec q\vec
r}\ket{\varphi^{\lambda}_l}\bra{\varphi^{\lambda'}_j} e^{-i\vec q\vec
r}\ket{\varphi^{\lambda'}_k}$ of the statically screened Coulomb potiential
$w_{\vec q}$ contain the single particle states $\ket{\varphi^\lambda_i}$ of
electrons and holes in the QD-WL system (see
Ref.~\cite{nielsen_many-body_2004}).
    The first two terms in Eqs.~(\ref{eq:we}),(\ref{eq:wh}) represent the
electrostatic (Hartree) interaction of the WL carriers with the excitonic
electron and hole. The last term in both equations is the exchange (Fock)
integral.

    Dissipative processes are included via Lindblad terms ${\cal L}_x(\rho) =
\frac{\gamma_x}{2}\left(2\, x \rho\, x^{\dagger} - x^{\dagger}x \rho -\rho\,
x^{\dagger}x \right) \,$ that involve transitions associated with the system
operator $x$, taking place at a rate $\gamma_x$. We consider cavity losses,
nonradiative decay of the QD-exciton and incoherent excitation with the
corresponding rates $\gamma_b =: \kappa$,  $\gamma_{X} =: \Gamma$ and
$\gamma_{X^{\dagger}} =: P$, respectively.

    The idea of the Schrieffer-Wolff approach (SWA) is to use a unitary
transform: $H' = e^S\,H\, e^{-S} = H + \left[S,H \right] + \frac{1}{2}\left[S,
\left[S,H\right]\right] + \dots \, ,$ with $S$ selected such that the JC
Hamiltonian is formally eliminated. This can be achieved by choosing $ S =
-\frac{g}{\Delta}\left(b^\dagger X - b\, X^\dagger \right)$\,, which shows that
the generator $S$ is of first order in a series expansion in the small parameter
$g/\Delta$, where $\Delta= \varepsilon_X - \omega$ is the exciton-cavity
detuning. Typically $g$ is of the order of $0.1$\,meV and for large detunings
($\Delta \sim $ $1-10$\,meV) a perturbative approach is justified. Neglecting
terms of higher order than $(g/\Delta)^2$, and $(g/\Delta)\,
W$\,~\cite{hohenester_cavity_2010}, one is lead to an effective JC interaction
Hamiltonian
    \begin{align}
        H'_\mathrm{int,SWA} = - \frac{g}{\Delta}\, W \,
(b^\dagger\,X+b\,X^\dagger)
    \end{align}
    that describes transitions between the QD exciton and the cavity photons,
assisted by the Coulomb interaction with WL carriers.

    When evaluating the QD exciton and photon dynamics under the influence of
$H'_\mathrm{int,SWA}$, the quasi-continuous nature of the delocalized WL-states
allows for treating them as a bath: We consider the fermionic WL reservoir
as being stationary and in thermal equilibrium. Following the standard
Born-Markov approach~\cite{breuer_theory_2007, roy_influence_2011} for the
system-bath interaction, we find new Lindblad terms, ${\cal L}_{b^\dagger X}$
and ${\cal L}_{b\,X^\dagger}$, with rates given by the Fourier transform of the
reservoir correlator $\braket{W (t)  W(0)} - \braket{W}^2$, taken at the energy
$\Delta$ lost in the transition corresponding to the exciton recombination. As a
result we obtain
    \begin{align}
        \gamma_{b^\dagger X} = 2 \pi\frac{g^2}{\Delta^2}\sum_{\lambda,\vec
k,\vec k'} \big |W^{\lambda}_{\vec k,\vec k'}\big|^2\, f^{\lambda}_{\vec
k}(1-f^{\lambda}_{\vec k'})\, \delta(\Delta + \varepsilon^{\lambda}_{\vec k\vec
k'}) \,.
        \label{eq:gamma_bTX}
    \end{align}
    Here $\varepsilon^{\lambda}_{\vec k\vec k'}=\varepsilon^\lambda_{\vec k} -
\varepsilon^\lambda_{\vec k'}$ and the occupancies $f^\lambda_{\vec
k}=\braket{\lambda^\dagger_{\vec k} \lambda^{\pda}_{\vec k}}$ are
Fermi functions describing the WL carrier population. Similarly,
$\gamma_{b\,X^\dagger}$ follows by changing $\Delta$ to $-\Delta$. The
first-order mean-field contribution of $H'_\mathrm{int,SWA}$ maintains the form
of the JC Hamiltonian, but with a renormalized coupling strength $\tilde g =
\frac{g}{\Delta}\,\braket{ W}$, where $\braket{W} = -\sum_{\lambda,\vec k}
W^{\lambda}_{s\vec k,s\vec k}\, f^\lambda_{\vec k}$\,~\footnote{The Hartree terms
are compensating each other, because they involve the $\vec q=0$ Fourier
component of the Coulomb potential $w_{\vec q}$, which is canceled due to global
charge neutrality.}. Consequently, the Schrieffer-Wolff procedure modifies the
coherent energy exchange between the exciton and the cavity and at the same time
gives rise to an incoherent exchange in the form of a new type of Lindblad
terms.
    \begin{figure}[t]
        \centering
        \includegraphics[width=0.49\textwidth]{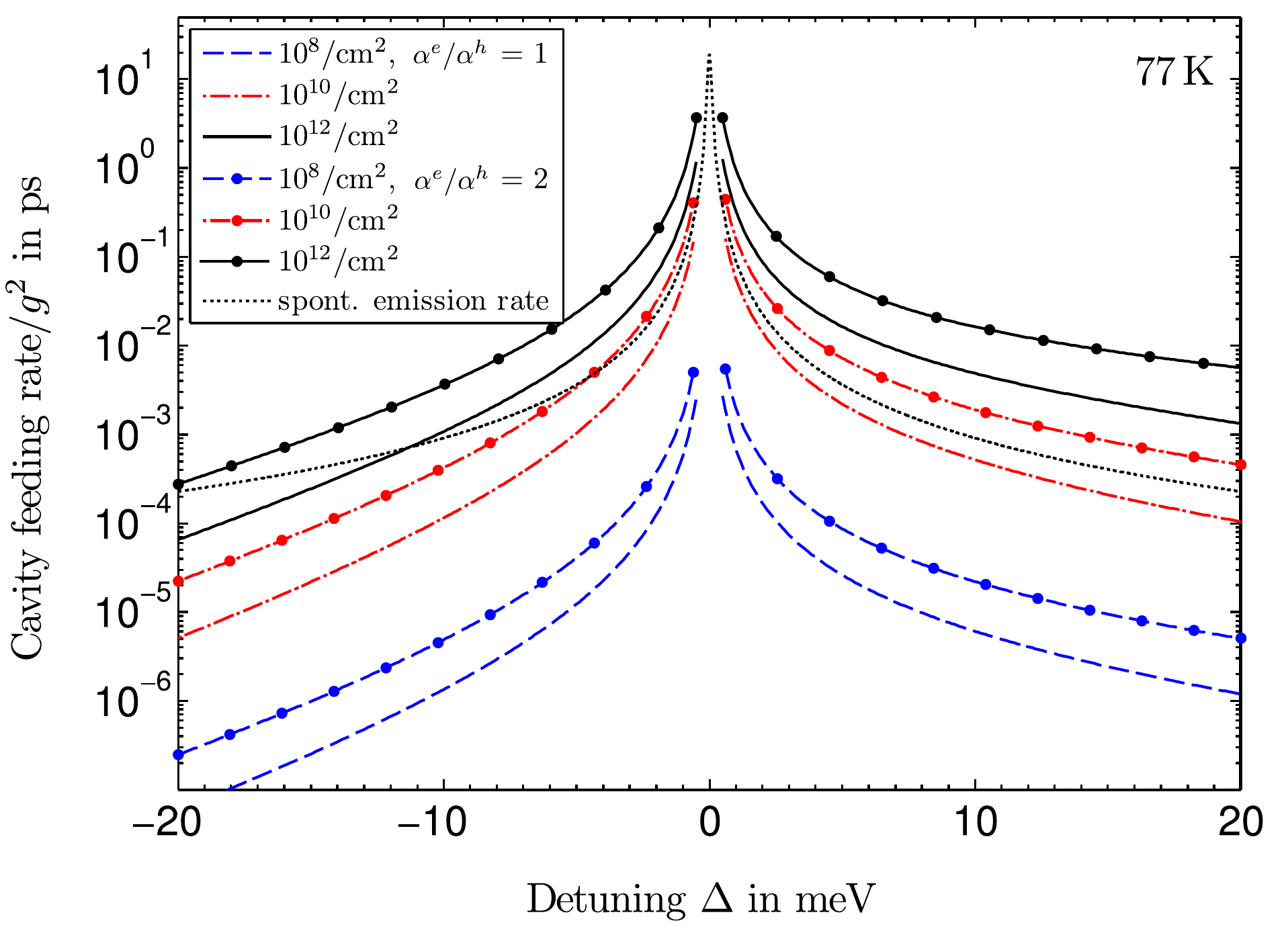}
        \caption{Cavity-feeding rate for a non-resonant QD coupled to the
continuum of WL-states at a temperature of $77$\,K obtained by using the SWA.
        We vary the WL carrier density $n_{\mathrm{WL}}$ from $10^8$ --
$10^{12}$/cm$^2$ and consider electron and hole envelopes that are equal (lines
without dots) or differ by a factor of two (lines with dots). In all
calculations we use $\alpha^h = 0.15/$nm. For comparison the spontaneous
emission rate is shown (dotted line) for typical parameters ($\kappa=0.1/$ps,
$\Gamma=0.01/$ps, $P=0.1/$ps). Note that the rates are normalized to the square
of the light-matter coupling strength.}
        \label{fig:SW_rate}
    \end{figure}

    In Fig.~\ref{fig:SW_rate} we address the dependence of the Coulomb assisted
cavity feeding rate $\gamma_{b^\dagger X}$ on the detuning $\Delta$ between
exciton and cavity mode for different carrier densities $n_{\mathrm{WL}}$ in the
delocalized WL states at a temperature of $77$K. In the calculation, we have
used InGaAs parameters~\cite{steinhoff_treatment_2012} and assume a flat
lens-shaped QD on a WL. For the QD wave functions of the energetically lowest
confined states $\varphi^\lambda_s(\vec r)$ Gaussian functions with standard
deviation $\alpha^\lambda$ are applied for the in-plane motion and in growth
direction the solution of a finite height potential well is assumed. For the WL
states $\varphi^\lambda_{\vec k}(\vec r)$ orthogonalized plane waves are
used~\cite{bockelmann_electron_1992}. With increasing WL-carrier density
additional scattering channels can compensate for the energetic mismatch
$\Delta$ between exciton and cavity, which leads to an increasing feeding rate.
For comparison, we show 
the spontaneous emission rate $R = 4g^2(\kappa+\Gamma+P)/[(\kappa+\Gamma+P)^2 +
(2\Delta)^2]$ caused by the JC coupling alone for typical QD-cavity parameter.
At low carrier density Coulomb assisted cavity feeding is negligible in
comparison to the spontaneous emission rate, which is in agreement with previous
experiments performed under low excitation
condition~\cite{ates_non-resonant_2009, press_photon_2007}, in which only phonon
signatures were found. However, for sufficiently high carrier densities
($n_\mathrm{WL}>10^{10}/$cm$^2$) Coulomb assisted processes prevail even at
large detuning and lead to a significant cavity feeding that is one order of
magnitude stronger in comparison to the JC coupling alone. Secondly, there is a
pronounced asymmetry between positive and negative detuning for low temperature.
This is expected, since any thermal bath favors the process which lowers the
system energy, all the more so at low temperatures. 

Finally, we find a significant 
reduction of the off-resonant cavity feeding, if the wave functions for
electrons and holes are similar (lines without dots). The reason is a large
degree of compensation between the electrostatic (Hartree) Coulomb integrals
contributing to Eq.~\eqref{eq:gamma_bTX}.
For identical electron and hole wave functions the exciton is not only globally
but also \textit{locally} neutral and there is no electrostatic interaction
between the exciton and the WL carriers. Classically, the system and the bath
become uncoupled, only the exchange interaction is left. This is made obvious
in Eqs~(\ref{eq:we}),(\ref{eq:wh}), where the integrals become $e,h$-independent
and the direct terms cancel out. Thus only the typically much smaller exchange
(Fock) integrals remain. This points to an intrinsic difference between
interband cavity assisted feeding and intraband scattering processes: In the
latter case electrons and holes can scatter independently, while in the former
case the emission of a photon requires the presence of an exciton, i.e.,
a fully correlated electron-hole pair. As a consequence any formalism describing
the off-resonant cavity feeding, which relies on an interaction of Coulombian
origin, like QD-WL Auger interaction, or Fr\"ohlich interaction of QD carriers
with LO phonons, must obey the following local neutrality condition: \textit{for
locally neutral excitons and discarding the exchange terms the off-resonant
process should vanish exactly}. 

    So far, we have presented results based on an approximate, perturbative  
diagonalization of the JC interaction part of the Hamiltonian. 
As an alternative approach, we introduce a non-perturbative treatment of the
Coulomb interaction between QD excitons and WL carriers by using a suitable
unitary transform.  
The idea is in many ways similar to the polaron transform of the IBM. 
The presence of the exciton generates an external field to which the lattice
ions react by displacements of their oscillation centers. The polaronic
transform connects the distorted lattice with the original one. Similarly, in
the present case the exciton perturbs the WL-carrier system by an external field
term $W$. The source of this field is localized around the QD, which thus acts
as a scattering center for the continuum of extended WL states. The associated
$\mathcal S$-matrix is the unitary transform that connects $h_X$ and $h_0$ (see
e.g.~\cite{sitenko_scattering_2012}). Even though the Coulomb interaction
with the WL states cannot be treated exactly any more, it lends itself to a
diagrammatic expansion. We term the method ``scattering potential approach''
(SPA).

    Specifically, one has
    \begin{align}
        h_0 &= \mathcal S(-\infty,0) \, h_X \, \mathcal S(0,-\infty) \, ,
        \label{eq:hh0}
    \end{align}
    with $\mathcal S$ generated by the scattering potential $W$
    \begin{align}
        \begin{split}
            \mathcal S(t_1, t_2) &= \mathcal S(t_1-t_2) = \mathcal
S^\dagger(t_2,t_1)\\
            &=\mathcal{T} \exp \left[ -i \int_{t_2}^{t_1} \, W(t)\,\text{d}t
\right] \,, \quad t_1>t_2 \, ,
        \end{split}
    \end{align}
    assuming that the requirements of the scattering theory are fulfilled
~\cite{sitenko_scattering_2012}. $\mathcal{T}$ is the time ordering operator and
the interaction representation of the perturbation $W(t)$ with respect to $h_0$
is used. We formally eliminate the QD-WL Coulomb interaction part from the
Hamiltonian with a unitary transform $U = 1-X^\dagger X + X^\dagger X \,\mathcal
S(0,-\infty) \, ,$ which switches on the action of the $\mathcal S$-matrix only
when the exciton is present. The interaction with the WL appears now in the
light-matter coupling term by the presence of the $\mathcal S$-matrix
    \begin{align}
        H'_{\mathrm{int,SPA}} = g\left[b^\dagger X\,\mathcal S(0,-\infty) +
b\,X^\dagger \mathcal S(-\infty,0) \right] \, .
    \end{align}
    The standard Born-Markov treatment of the system-reservoir interaction with
respect to $H'_{\mathrm{int,SPA}}$ leads to Lindblad terms of the form ${\cal
L}_{b^\dagger X}$ and ${\cal L}_{b X^\dagger}$. For the off-resonant cavity
feeding rate $\gamma_{b^\dagger X}$, we obtain
    \begin{align}
        \gamma_{b^\dagger X} = 2 g^2 e^{\beta\Delta} \Re\int_{0}^\infty e^{-i
\Delta t}\,\braket{\mathcal S(t,0)}\text{d} t \, ,
        \label{eq:gamma}
    \end{align}
    with the inverse temperature $\beta=1/(k_\mathrm{B} T)$.
    The evaluation of $\braket{ \mathcal S(t,0) }$  can be done using the linked
cluster (cumulant) expansion~\cite{mahan_many_2000, abrikosov_methods_1975}.
Details about the derivation are included in Appendix~\ref{sec:SPA}.
    \begin{figure}[t]
        \centering
        \includegraphics[width=\columnwidth]{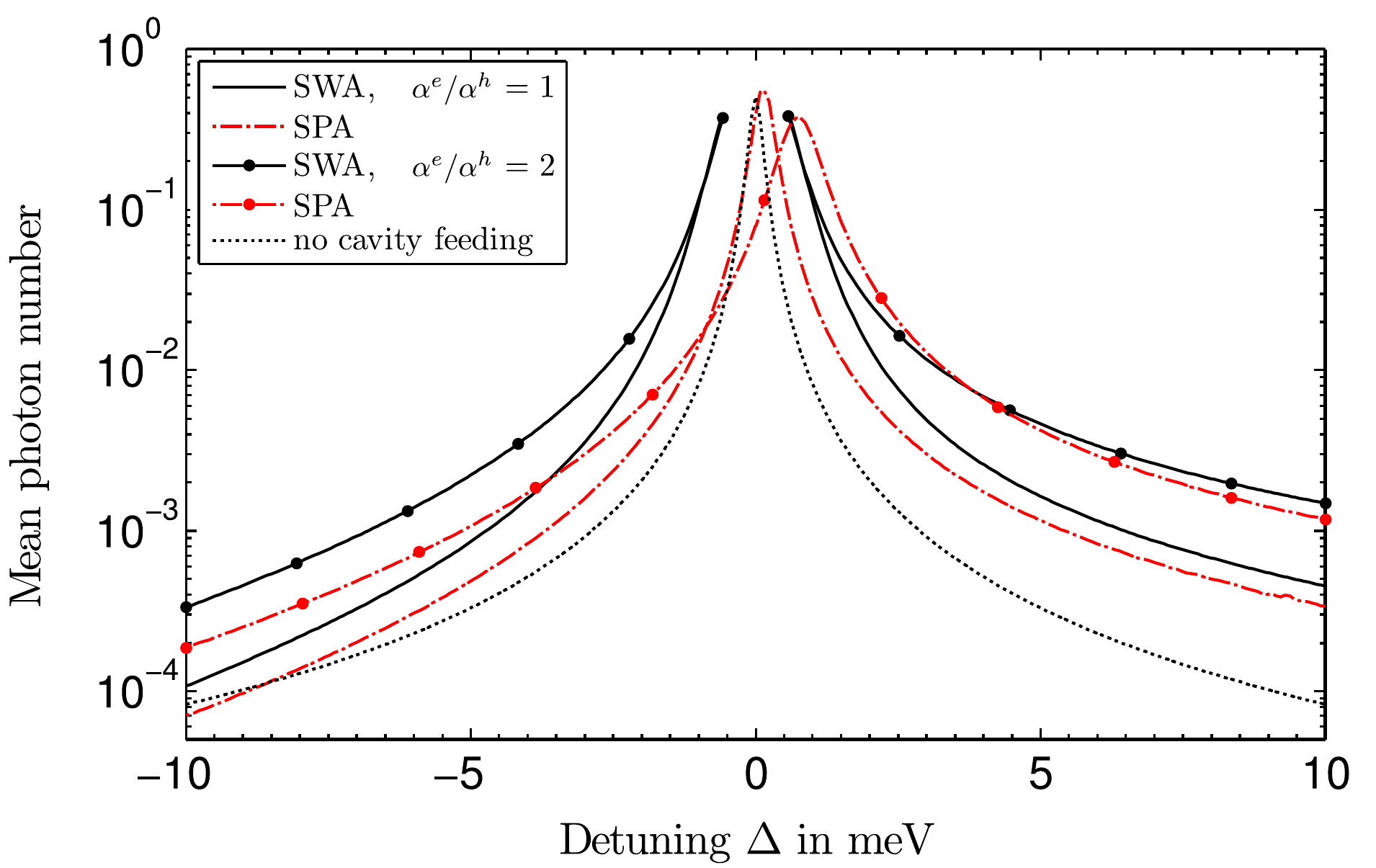}
        \caption{Steady state cavity mean photon number generated by an
off-resonant QD mediated by the continuum of WL states for a carrier density of
$10^{12}/$cm$^2$ by using the SWA (solid) and the SPA (dashed dotted) for a
temperature of $77$\,K. Additionally, results for non-equal envelopes (lines
with dots) and pure JC coupling (dotted) are shown. Typical parameters are used
($g=0.1/$ps, $\kappa=0.1/$ps, $\Gamma=0.01/$ps), $P=0.1/$ps.}
        \label{fig:mean_cmp}
    \end{figure}
    \begin{figure}[t]
        \centering
        \includegraphics[width=\columnwidth]{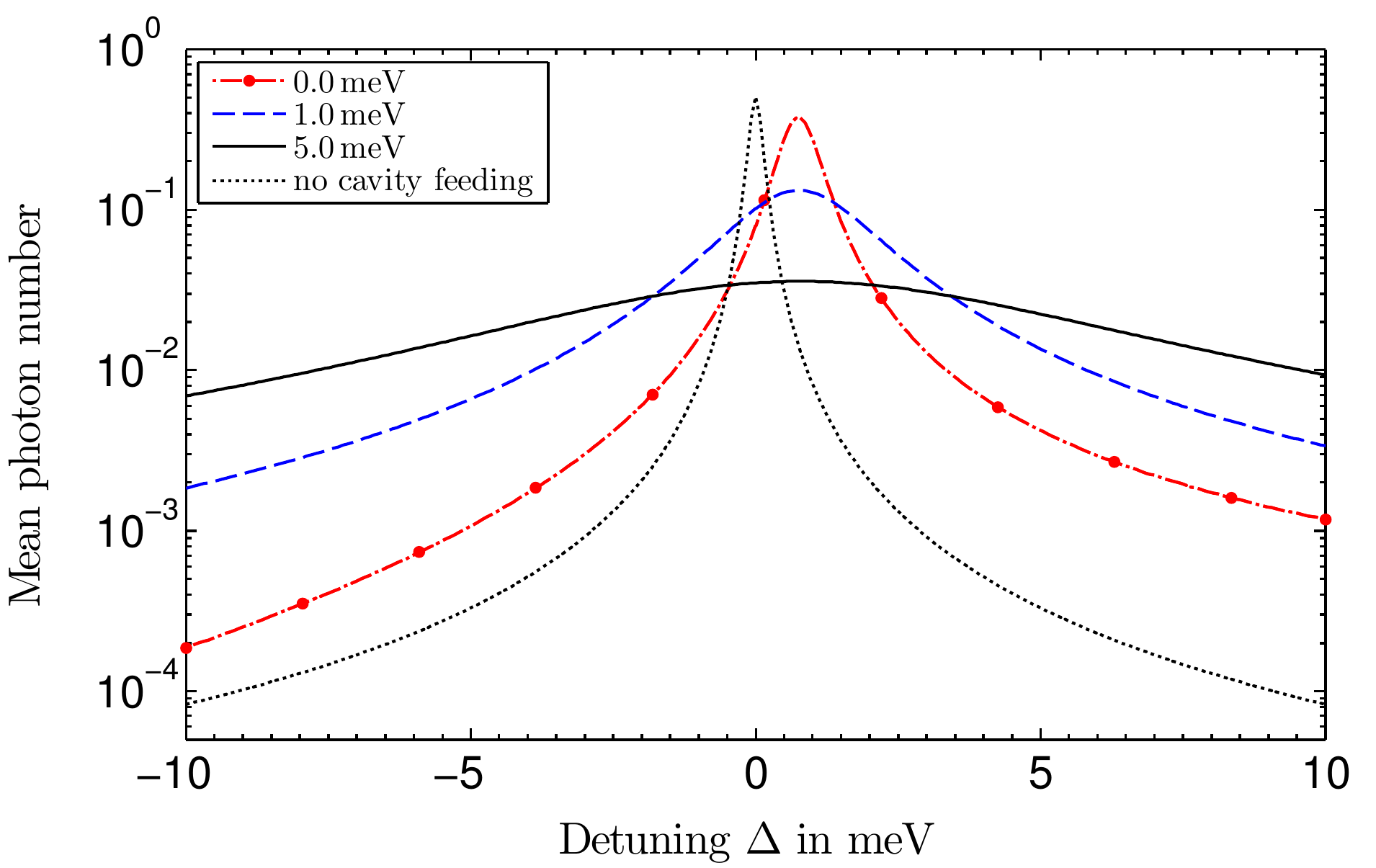}
        \caption{Influence of artificial renormalization/broadening of the SPA
result, as a consequence of uncorrelated electrons and holes. Shown is the
steady state mean photon number for a carrier density of $10^{12}/$cm$^2$, a
temperature of $77$\,K and non-equal envelopes $\alpha^e/\alpha^h = 2$. Without
broadening the dash dotted line corresponds to the result in
Fig.~\protect\ref{fig:mean_cmp}.}
        \label{fig:mean_ren}
    \end{figure}

    In Fig.~\ref{fig:mean_cmp} we show the steady-state mean photon number for a
typical QD-cavity system and compare the results obtained by the SWA (black
line) and the SPA (red dash dotted line) for a high WL carrier density of
$n_{\mathrm{WL}} = 10^{12}$/cm$^2$, where we expect a pronounced off-resonant
cavity feeding. To this end, we solve the time evolution of the system density
operator by the von-Neumann Lindblad equation $\frac{\partial}{\partial t}\rho =
-i \left[H_\mathrm{sys}, \rho \right] + \sum_x {\cal L}_x(\rho) \,$ and include
the developed cavity feeding Lindblad terms ${\cal L}_{b^\dagger X}$ and ${\cal
L}_{b\,X^\dagger}$. We find a good qualitative agreement between both methods
for large detunings. As an advantage, the non-perturbative SPA allows to extend
the results to small detuning values and includes additional WL-induced
renormalizations.

    The compensation effect between electron and hole potentials that we
discussed for the SWA also holds for the SPA, as the perturbation acting on the
WL is the same one-particle operator $W$. Thus, electrons and holes are not
contributing independently, and this results in a reduced cavity mean photon
number for a locally neutral QD (lines without dots). A calculation beyond the
second-order Born-Markov approach for the system-bath treatment, e.g. by using
quantum kinetic methods, would involve quasi-particle properties with complex
spectral structure instead of the free-particle energies. In the simplest
approximation this leads to a Lorentzian broadening of the electron and hole
energy spectrum. However, the renormalization of the exciton energy should also
take into account the correlations between the two constituent particles and
especially reflect the reduced QD-WL Coulomb interaction for a locally neutral
exciton. In this case the exciton renormalization should be considerably smaller
in comparison 
to the sum of independently broadened electron and hole energies. In
Fig.~\ref{fig:mean_ren} we show how artificially uncorrelated electrons and
holes would influence the cavity feeding. To this end we introduce an additional
renormalization to the QD exciton energy in Eq.~\eqref{eq:gamma} by replacing
$\varepsilon_X \rightarrow \varepsilon_X + i\gamma$, where
$\gamma=\gamma_e+\gamma_h$ is the sum of the independently broadened electron
and hole energies. It is seen that by increasing the total broadening $\gamma$
the off-resonant cavity feeding gains substantially. Being an artifact, this
result underscores the need for the correct treatment of correlations between
the electron and hole, which reduce the efficiency of interband transitions
assisted by wetting layer carriers.

In conclusion, for the off-resonant coupling between QD emitters and optical cavity modes, 
the Auger-like recombination assisted by WL carriers has been identified as a possible 
cavity feeding channel. It coexists with other processes like phonon-assisted
recombination, which is only efficient for small detuning, or multi-exciton effects, which require QDs
with several shells. 
For large carrier densities in the delocalized states, we have demonstrated efficient coupling at large detunings up to 10\,meV
for a QD with only one confined shell. 
The off-resonant coupling of QD emitters has profound implications for the description and characterization of QD microcavity lasers, where a small number of close to resonant emitters are accompanied by many far-detuned emitters. Their background emission can enhance the photon production rate sufficiently to result in a reduced laser threshold and modified photon statistics. Finally, we point out that the described assisted interband scattering processes, while significant, are found to be much smaller in comparison to the intraband-relaxation processes. The reason lies in the correlation between electron and hole, that is required for the interband recombination, but not for the intraband scattering process. 

    \begin{acknowledgments}
        The authors acknowledge financial support from the Deutsche
Forschungsgemeinschaft (DFG) and the Bundesministerium f\"ur Bildung und
Forschung (BMBF).
    \end{acknowledgments}

    \appendix
\section{The exciton-WL interaction}
\label{sec:exc_WL}
    The Coulomb interaction between the QD and the WL carriers is expressed as 
    \begin{align}    
       \begin{split}    
       &W = \\    
       &e^\dagger e \sum_{\vec k,\vec k'} \left[  
       W^{ee}_{s\vec k,\vec k's}e^\dagger_{\vec k} e^{\pda}_{\vec k'} -
       W^{eh}_{s\vec k,\vec k's}h^\dagger_{\vec k} h^{\pda}_{\vec k'} -
       W^{ee}_{s\vec k,s\vec k'}e^\dagger_{\vec k} e^{\pda}_{\vec k'}
        \right] + \\
       &h^\dagger h \sum_{\vec k,\vec k'} \left[  
       W^{hh}_{s\vec k,\vec k's}h^\dagger_{\vec k} h^{\pda}_{\vec k'} -
       W^{he}_{s\vec k,\vec k's}e^\dagger_{\vec k} e^{\pda}_{\vec k'} -
       W^{hh}_{s\vec k,s\vec k'}h^\dagger_{\vec k} h^{\pda}_{\vec k'}
        \right] \, .        
       \end{split}
      \label{eq:x_wl}
    \end{align}
The subscripts $\vec k, \vec k'$ describe WL continuum states, while electron $e,
e^\dagger$ and hole $h, h^\dagger$ operators without subscripts correspond to
the QD-localized $s$-states. The two lines in Eq.~\eqref{eq:x_wl} define the contribution from the
QD electron and hole respectively. 

The Hilbert space of our problem is limited to neutral states. In other words
the QD electron and hole are either both absent or both present, which prohibits an approach where Lindblad terms are 
derive from each of the two lines of Eq.~(\ref{eq:x_wl}) separately, because the excitonic pair is highly
correlated. This is taken into account by having $e^\dagger e = h^\dagger h =
X^\dagger X $, where $X=he$, and leads to the exciton-WL Hamiltonian considered
in the paper, which accumulates the terms of both lines. The ensuing
compensation between the fields created by the exciton carriers, and its
consequences, are discussed in the paper.

\section{Scattering potential approach}
\label{sec:SPA}
    We formally eliminate the QD-WL Coulomb interaction part from the Hamiltonian 
    \begin{align}
        \begin{split}
            H =&\,\varepsilon_X X^\dagger X + \omega b^\dagger b + g(b^\dagger X + b X^\dagger)\\
            &+(1 - X^\dagger X)h_0 + X^\dagger X h_X\,,
            \label{eq:hamiltonian}
        \end{split}
    \end{align}
    using the unitary transform
    \begin{align}
        U = 1-X^\dagger X + X^\dagger X \,\mathcal S(0,-\infty) \, ,
    \end{align}
        with the properties that $X' = U^\dagger X U = X\, \mathcal S(0,-\infty)$, ($X^\dagger)'  = \mathcal S(-\infty,0)\,X^\dagger $, while $X^\dagger X$ remains invariant. The important changes concern the WL part of the Hamiltonian of Eq.~(\ref{eq:hamiltonian})
    \begin{align}
            U^\dagger & \left[(1-X^\dagger X)\, h_0 + X^\dagger X \, h_X \right] U =(1-X^\dagger X)\, h_0 \nonumber \\
            &+ X^\dagger X\, \mathcal S(-\infty,0)\,h_X\, \mathcal S(0,-\infty) = h_0 \, ,
    \end{align}
    from which the interaction is now removed. The same role is played by the polaron transform in the IBM. 
    
    For the transformed light-matter coupling Hamiltonian we obtain
    \begin{align}
        H'_{\mathrm{int,SPA}} = g\left[b^\dagger X\,\mathcal S(0,-\infty) + b\,X^\dagger \mathcal S(-\infty,0) \right] \, .
    \end{align}
        The standard Born-Markov treatment of the system-reservoir interaction with respect to $H'_{\mathrm{int,SPA}}$ leads to Lindblad terms of the form ${\cal L}_{b^\dagger X}$ and ${\cal L}_{b X^\dagger}$, where the corresponding rates $\gamma_{b^\dagger X}$ and $\gamma_{b X^\dagger}$ are 
        \begin{align}
            \begin{split}
                \gamma_{b^\dagger X} = & g^2\int_{-\infty}^\infty e^{\,i \Delta
t}\, \braket{ e^{ih_0t}
                \mathcal S(-\infty,0) e^{-ih_0t} \mathcal S(0,-\infty) } \text{d} t \\
                \gamma_{bX^\dagger} = &  g^2\int_{-\infty}^\infty e^{-i \Delta
t} \,\braket{ e^{ih_0t}
                \mathcal S(0,-\infty) e^{-ih_0t} \mathcal S(-\infty,0) } \text{d} t \, ,
            \end{split}
        \end{align}  
with the averages taken over the thermal equilibrium of $h_0$. The rates obey the Kubo-Martin-Schwinger (KMS) relation 
    \begin{align}
        \gamma_{b^\dagger X} = e^{\beta\Delta} \,\gamma_{b X^\dagger}
    \end{align}
    with the inverse temperature $\beta=1/(k_\mathrm{B} T)$. Therefore it is
sufficient to compute only one of these rates. The simplest one is  $\gamma_{b
X^\dagger}$. The correlator $\braket{e^{ih_0 t}\mathcal
S(0,-\infty)e^{-ih_0t}\mathcal S(-\infty,0)}$ can be rewritten as
$\braket{\mathcal S(t,0)}$ by using succesively $\mathcal
S(0,-\infty)e^{-ih_0t} =  e^{-ih_Xt}\mathcal S(0,-\infty)$, in
accordance with the unitary equivalence between $h_X$ and $h_0$ 
    \begin{align}
       \mathcal S(0,-\infty) \, h_0 = h_X \, \mathcal S(0,-\infty) \,,  
    \end{align}
    $e^{ih_0t} e^{-ih_Xt}=\mathcal S(t,0)$, and the semigroup property of the $\mathcal S$-matrix. 
    For the off-resonant cavity feeding rate $\gamma_{b^\dagger X}$, we obtain
    \begin{align}
        \gamma_{b^\dagger X} = 2 g^2 e^{\beta\Delta} \Re\int_{0}^\infty e^{-i
\Delta t}\,\braket{\mathcal S(t,0)}\text{d} t \, ,
        \label{eq:gamma}
    \end{align}
where we have also used that $\braket{\mathcal S(-t,0)}=\braket{\mathcal
S(0,t)}=\braket{\mathcal S(t,0)}^* $. 
    
    The evaluation of $\braket{ \mathcal S(t,0) }$  can be done using the linked
cluster (cumulant) expansion. This is expressed as $\braket{ \mathcal S(t,0) } =
\exp[\Phi(t)] \, ,$ where $\Phi(t) = \sum_n L_n(t)$ is the sum over all
connected diagrams $L_n$ with no external points, having $n$ internal ones and
carrying a prefactor $1/n$. The internal points run in time from 0 to $t$.
    \begin{figure}[t]
        \centering
        \includegraphics[width=\columnwidth]{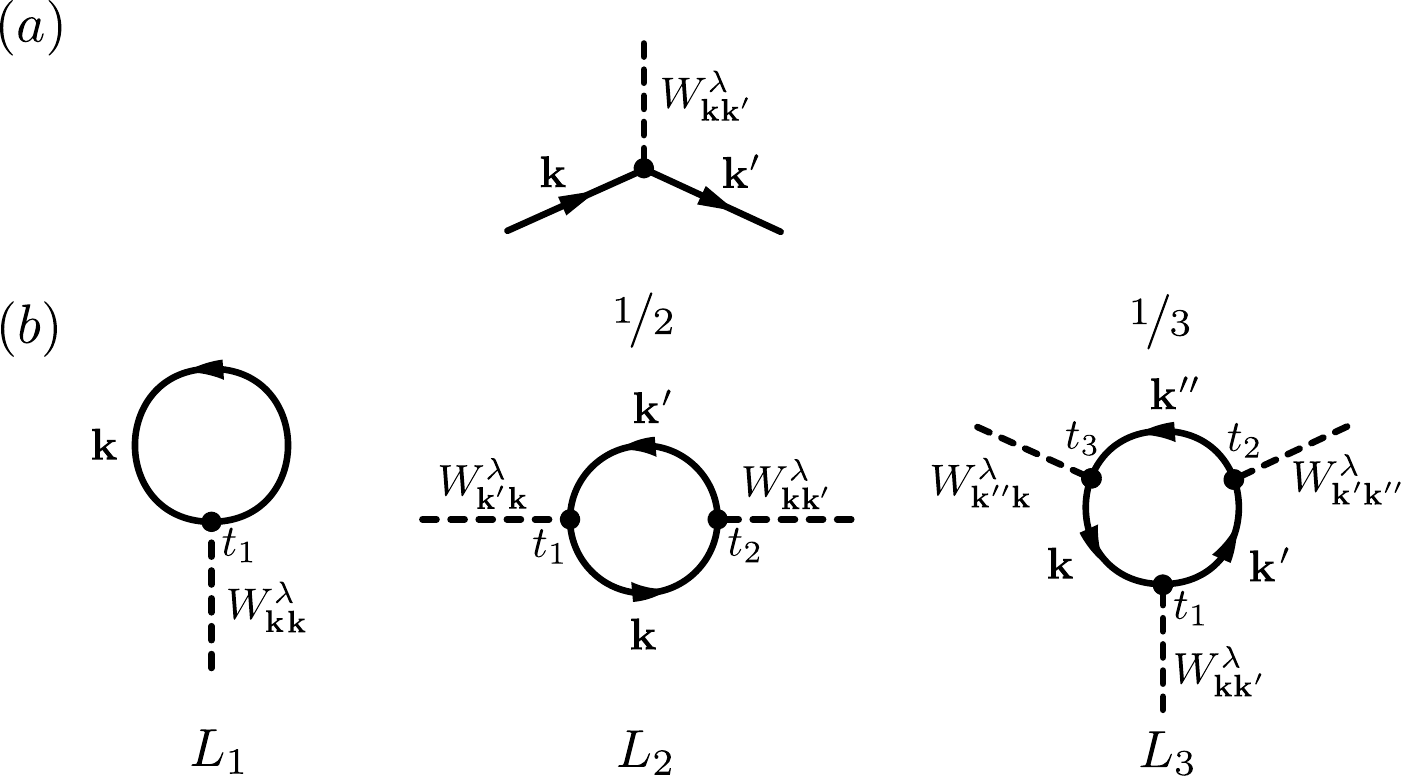}
        \caption{(a) Elementary interaction vertex. (b) First three connected diagrams $L_1$, $L_2$ and $L_3$ of the linked cluster expansion in the evaluation of the thermal average of the $S$-matrix $\braket{S(t,0)}$.}
        \label{fig:diagram}
    \end{figure}
    Here, the interaction is an external potential, not a many-body one, and the elementary interaction vertex, $W^{\lambda}_{\vec k,\vec k'}$, is represented in Fig.~\ref{fig:diagram}(a). The first terms in the linked cluster expansion are shown in Fig.~\ref{fig:diagram}(b). In what follows we restrict the calculation to the first two diagrams. Note that in the case of interaction with phonons the result is exact, because only the corresponding $L_2$ diagram is present in the IBM theory. One has
    \begin{align}
        \begin{split}
            L_1(t) =& - \sum_{\lambda,\vec k} \int_0^t \text{d}t_1 W^{\lambda}_{\vec k,\vec k}\, G^0_{\lambda \vec k}(t_1,t_1^+) \\
            L_2 (t)=& -\frac{1}{2} \sum_{\lambda, \vec k,\vec k'} \Big|W^{\lambda}_{\vec k,\vec k'}\Big|^2 \\
            &\times\int_0^t \text{d}t_1\int_0^t\text{d}t_2 \,G^0_{\lambda \vec k}(t_1,t_2) \, G^0_{\lambda \vec k'}(t_2,t_1)\,,
        \end{split}
        \label{eq:diag_uneval}
    \end{align}
    where $G^0_{\lambda \vec k}$ is the free Green's function for the WL state $\lambda \vec k$ and $t_1^+$ is infinitesimally later than $t_1$. Evaluating the time integration finally leads to
    \begin{align}
        \begin{split}
        L_1(t) =&- i\sum_{\lambda,\vec k} W^{\lambda}_{\vec k,\vec k}\, f^\lambda_{\vec k}\,\, t\\
        L_2(t) =&\sum_{\lambda,\vec k,\vec k'} \Big| W^{\lambda}_{\vec k,\vec k'}\Big|^2 (1-f^{\lambda}_{\vec k})f^{\lambda}_{\vec k'}\,\\
                &\times \Big( e^{-i\,\varepsilon^{\lambda}_{\vec k\vec k'}\,t} -1 +i\,\varepsilon^{\lambda}_{\vec k\vec k'}\,t \Big)\Big(\varepsilon^{\lambda}_{\vec k\vec k'}\Big)^{-2}  \, .
        \end{split}
        \label{eq:diag_eval}
    \end{align}
    $L_1$ is purely imaginary and amounts to a shift of the excitonic energy, which we include in the detuning $\Delta$. Further renormalizations occur due to the imaginary part of the $L_2$ term. The real part of the latter shows a linear long-time asymptotics, according to $[\cos(\varepsilon t)-1]/\varepsilon^2 = -t \sin^2(\varepsilon t/2)/(\varepsilon^2t/2)$ which for large times behaves like $-\pi\delta(\varepsilon) t$. This gives rise to an exponential decay of $\braket{ \mathcal S(t,0) }$ as $t \to \infty$, which implies a vanishing first-order mean-field contribution of the system-bath interaction $H'_{\mathrm{int,SPA}}$. An exponential behavior of $\braket{\mathcal S(t,0)}$ suggests a slow (Lorentzian) decay of its Fourier transform which would entail, according to Eq.~(\ref{eq:gamma}), a divergent rate $\gamma_{b^\dagger X}$ in the limit of large $\Delta$. This is prevented by the quadratic behavior of $L_2$, and the corresponding Gaussian shape of $\braket{\mathcal S(t,0)}$
  in the low-time regime.

\end{document}